\documentclass[11pt]{article}
\usepackage[preprint]{acl}

\usepackage{times}
\usepackage{latexsym}
\usepackage[T1]{fontenc}
\usepackage{fvextra}

\usepackage[utf8]{inputenc}
\usepackage{hyperref}
\usepackage{microtype}
\usepackage{inconsolata}
\usepackage{graphicx}
\usepackage{booktabs}

\usepackage[textsize=tiny]{todonotes}

\title{LinkNav: Surfacing Interconnected Information in Scientific Articles}

\author{Sebastian Joseph$^{1}$\ \ \ \
Jennifer Healey$^{2}$\ \ \ \ 
Junyi Jessy Li$^{1}$\ \ \ \ 
Ani Nenkova$^{2}$
\\
$^1$The University of Texas at Austin,
$^2$Adobe Research \\
{\small \tt \{sebaj,jessy\}@utexas.edu, \{jehealey,nenkova\}@adobe.com} 
}  

\begin{document}
\maketitle
\begin{abstract}

We present LinkNav\footnote{LinkNav link: \href{https://linknav.netlify.app}{linknav.netlify.app}}, 
an enhanced experience for reading academic papers which makes explicit connections between related but non-adjacent passages.  To create the experience, we instruct a language model to generate questions that may arise while reading a passage and then search for answer passages elsewhere in the document, forming intra-document connections when answers are found. We confirm that these building blocks work well to power the experience, with an answer detection pipeline that works with high precision, resulting in a reasonable number of connections being made for a document.
 On a dataset of academic papers, we find that connected passages are on average ten segments away from each other, making explicit connections that a reader may have otherwise missed.  

\end{abstract}

\section{Introduction}

Academic papers present hypotheses, experiments, mathematical derivations and evaluative statements about a complex body of work that have to be presented in linear form. 
In this complex landscape, it is normal to have semantic connections between segments of writing that are not adjacent in their linear presentation and that may be missed even by an attentive reader. A non-linear reading order of academic literature is common \cite{lo2023semanticreaderprojectaugmenting} with readers at different career stages approaching it differently \cite{hubbard2017perceptions}. 

The need to connect information is recognized by authors, who make explicit connections via pointers to sections and appendices. It also surfaces during informal academic talks, where it is commonplace for the audience to ask a question and the presenter to respond that the answer is to be discussed at a later point in the presentation.
In this paper we present analyses of questions drawn from peer reviews of scientific papers (cf. Section~\ref{PeerQA}); reviewers frequently ask questions that are already answered in a paper, further supporting the need and potential value of surfacing the connection between non-adjacent content.

\begin{figure}
    \centering
    \includegraphics[width=1\linewidth]{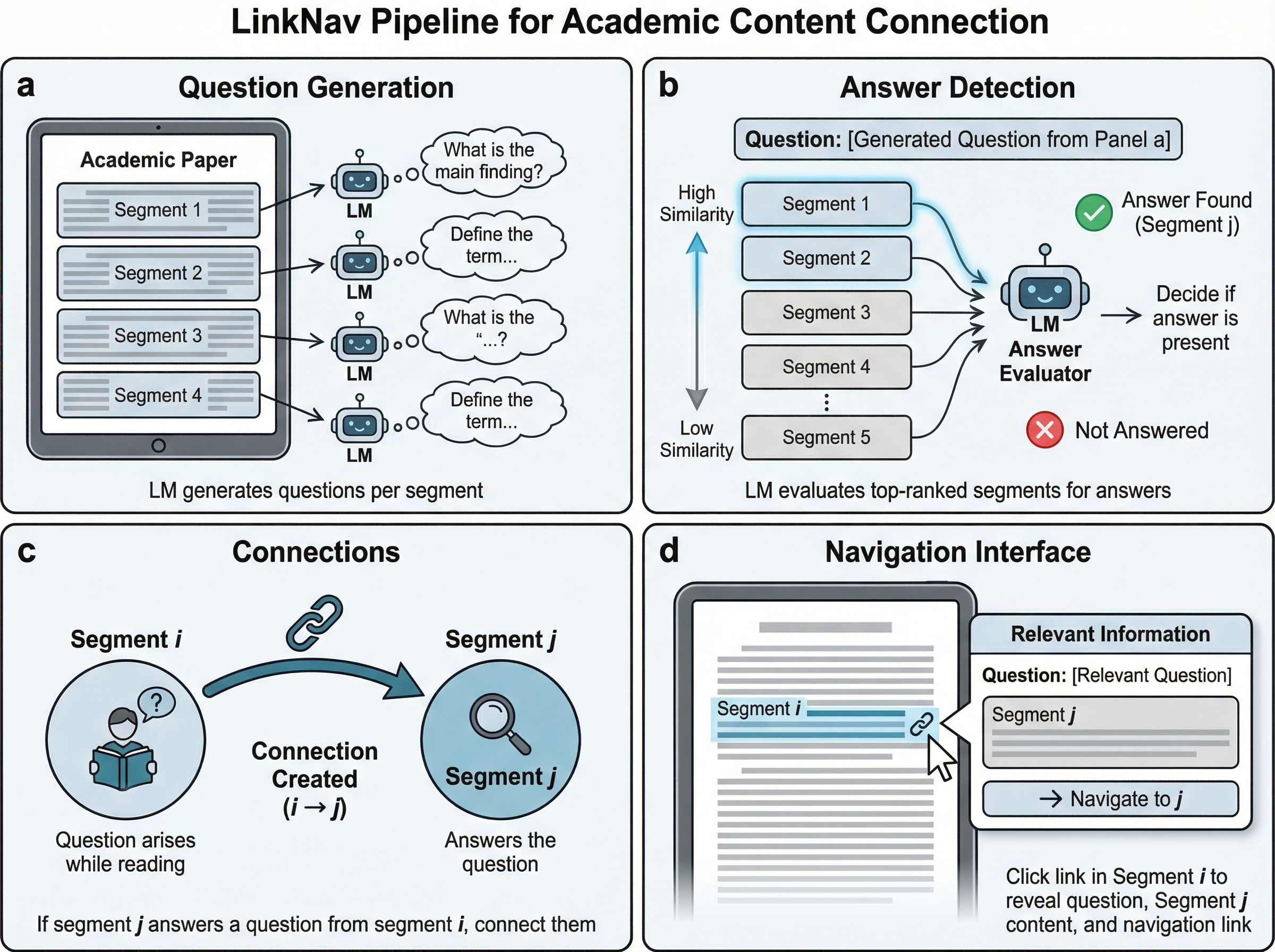}
    \caption{LinkNav pipeline and experience. \emph{(a)} A paper is broken down into segments and a language model generates questions for each segment $i$. \emph{(b)} Similarity is computed between the questions and all other segments. An LLM decides if one of the top five segments, $j$ is the answer. \emph{(c)} If so, $i$ and $j$ are considered connected. \emph{(d)} Connections are surfaced via links and a side panel.}
    \label{fig:LinkNavPipeline}
\end{figure}

\begin{figure*}
    \centering
    \includegraphics[width=1\linewidth]{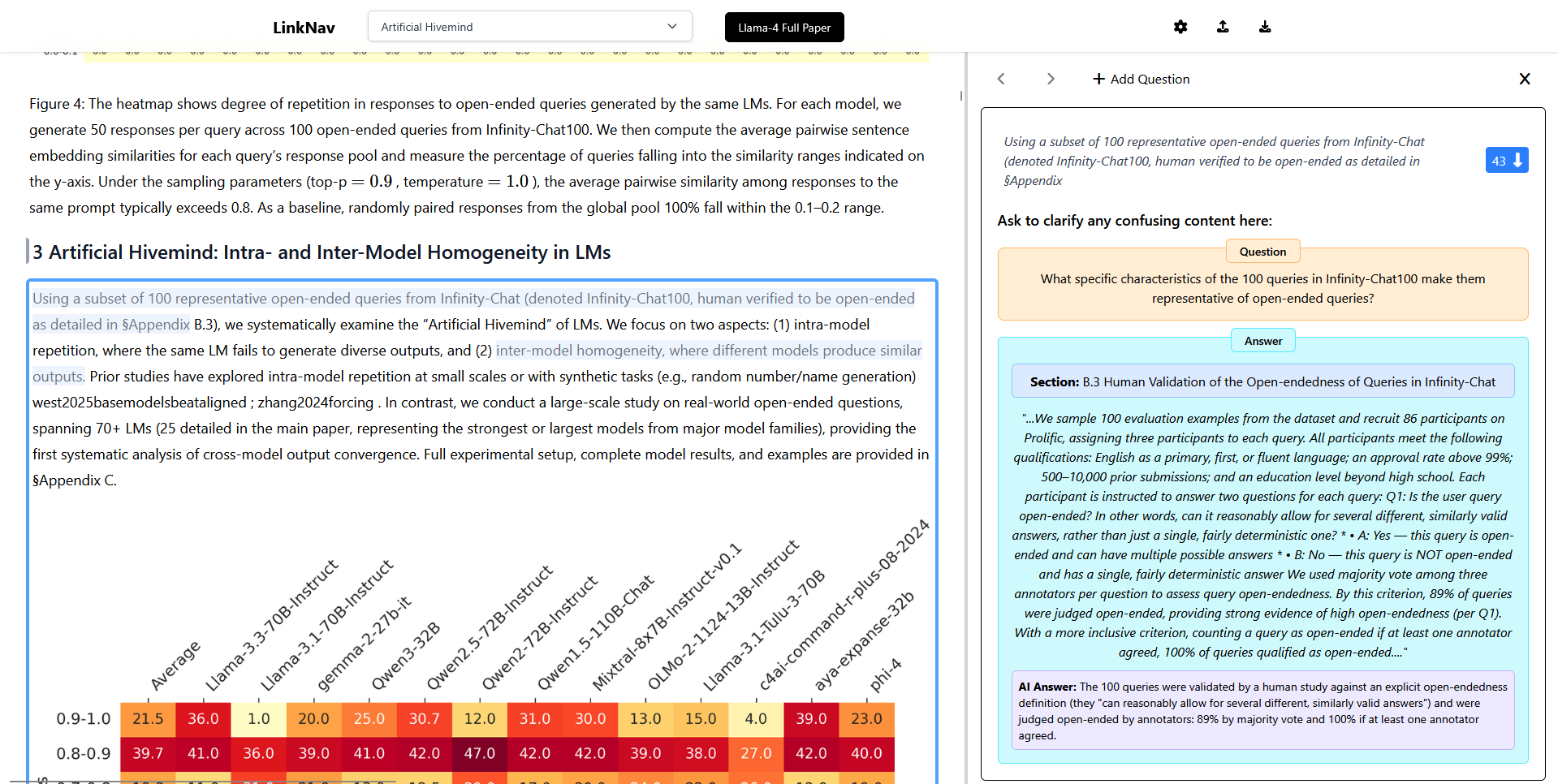}
    \caption{LinkNav surfaces links between different segments in papers. Text portions that likely triggered the question are displayed as links. Clicking on the link reveals a panel that displays the question, an extractive answer from the paper, a generative answer by a model presented with the question and the segment containing the answer. A link allows the reader to navigate to the segment where the answer is located. The distance between the originating segment and the answer segment is displayed. A quick glance will let the reader decide if they want to read the linked segment now, perhaps if it is very far away, or continue reading if the linked segment is nearby. }
    \label{fig:screenshot}
\end{figure*}

Figure \ref{fig:LinkNavPipeline}\footnote{Rendered with Gemini following a detailed author prompt.} schematically introduces LinkNav, an augmented reading interface which supplies connections even if the author did not make them explicit and displays the relevant content. 
Connections are established via the generation of inquisitive questions, an approach shown to relate to reader expectations \cite{wu-etal-2024-questions}.
LinkNav also provides a link to navigate to the related segment in the paper. Future user studies will show if people prefer the localization of information or navigation in the linear paper presentation; LinkNav supports 
both. Figure~\ref{fig:screenshot} shows a screenshot of the experience reading a paper in the interface, along with one of the questions and the extractive (author wording answer) and a generative answer (from a language model) provided with the question and the paper.

Using a subset of the PeerQA dataset~\cite{baumgärtner2025peerqascientificquestionanswering},
we analyze the methodologies used for this experience and the connections they create. Our analyses  show that LinkNav reveals explicit connections between distant areas of a document that a reader could have likely missed. 
Links connect for example information in appendices to the place where content is discussed, evaluative statements in the abstracts and the introduction to segments in the results and experiments sections of papers, giving the reader the ability to either skim more efficiently or do a close read with relevant details from the entire paper available where the reader is currently paying attention. 

\section{Related Work}

LinkNav is \emph{an augmented reading} experience driven by \emph{question generation}. We highlight the key connections with these lively research areas. 

\paragraph{Augmented reading}

Most closely related to our work is the Qlarify augmented reading experience \cite{fok2024qlarifyrecursivelyexpandableabstracts} for paper abstracts. There some entities are highlighted to indicate that they can be expanded. One contextual LLM-generated question about the entity is revealed when hovering over the link, while three static questions are always present for each entity---\emph{Define}, \emph{Expand}, \emph{Why}---which provide definition, concrete details about a concept and explanations about motivation. Readers can select text in the abstract and ask their own questions. The answers are linked to paper content supporting the answer, allowing readers to navigate to that part of the paper on demand.

A user study with doctoral students revealed that readers appreciated the cognitive load reduction provided by the suggested question. They asked their own questions only a quarter of the time.
A field study in which Qlarify was provided to conference attendees to browse the newly available proceedings confirmed the preference for pre-generated questions. In that setting, almost 90\% of interactions were via the three static questions. In a comparison with an interface that only allowed readers to ask their own questions instead of offering LLM generated questions, readers asked half as many questions. These findings inform our approach, emphasizing the importance of pre-surfacing interlinked content. 

ScholarPhi~\cite{10.1145/3411764.3445648} is an augmented reading experience for full academic papers. It solves only for the need for  just-in-time definitions of mathematical symbols used in equations and derivations. Like our work, it caters to a need occurring during reading of the full paper by bringing together information that is needed for understanding but is not adjacent in the linear presentation. DocVoyager \cite{10.1145/3706599.3719846} on the other hand expands the Qlarify idea of recursive invocation of information driven by questions, offering readers tailored non-linear reading experience. In contrast, LinkNav preserves the presentation flow of the author and leaves full control to the reader when they want to check the connection, without pre-supposing that it will guide the reader in their journey through the document.  Information cards for referenced papers, citing sentences from recently read papers, explanations and synthesis are also helpful to readers \cite{10.1145/3659096}.

\paragraph{Question Generation}

Suggested questions \cite{Huang2023-qu, Wang_Wei_Fan_Liu_Huang_2019, lin2024personasqpersonalizedsuggestedquestion} are now common in AI reading assistants. These are questions whose answers reveal noteworthy content in the document and that are made available to the reader to initiate and continue engagement with the assistant; they do not expose interconnected content. 

Similarly for medical documents \cite{paperplain}, ``key questions'' were used to guide readers to the answer, resulting in a non-linear reading order that is more efficient without loss in comprehension. These findings related to key questions demonstrate the potential of a question-generation approach. In  contrast to our work, key questions are akin to a fixed set of FAQs from clinicians, rather than generated dynamically based on context.

Questions most similar in function to the ones that underpin LinkNav are inquisitive questions, which may be evoked in the mind of a reader during comprehension \cite{ko-etal-2020-inquisitive,westera-etal-2020-ted,wu-etal-2024-questions}. Some of the inquisitive questions prepare the reader for content that will be presented later in the article. These questions typically have high \emph{salience} and readers have high expectation that they will be answered. \cite{wu-etal-2024-questions}. Other questions may not be answered because of gaps in cultural, topical or other background knowledge by the reader while the author assumes them as common ground. 

Earlier work on annotation of sentence specificity in news articles has revealed that different people find similar parts of the sentence to be underspecified \cite{LI16.930}. This prior work, along with \citet{westera-etal-2020-ted} and \citet{ko-etal-2022-discourse}, also found that often questions evoked in a sentence are answered in the immediately following sentence, as authors likely anticipate the question the reader may have or explicitly have phrased their content to make a piece of information salient in the reader's mind. A fair amount of questions were found to be answered \emph{before} the place where the question arose, with writing organized in a way that minimizes missing information. We incorporate these findings into LinkNav by pruning these author crafted links to content prior to a segment, or to the same or immediately following segment.  
We enrich the reading only with segments that occur after the place where the question arises. 

Generating questions that probe gaps in the knowledge conveyed by two texts or possessed by different people is useful in many scenarios such as measuring information loss during text simplification~\cite{trienes2024infolossqacharacterizingrecoveringinformation} or text editing~\cite{cole-etal-2023-diffqg}, education~\cite{rabin-etal-2023-covering} and automated paper reviewing~\cite{chang2025treereviewdynamictreequestions}. These scenarios are unrelated to augmented reading even though question generation is central ingredient in the solutions they provide.

\section{Reviewers May Benefit From LinkNav}
\label{PeerQA}

\paragraph{Data} We use the PeerQA dataset \cite{baumgärtner2025peerqascientificquestionanswering} for formative analyses here and for evaluation of LinkNav components later. PeerQA consists of questions contained in peer reviews of (mostly) machine learning and natural language processing venues. For part of the PeerQA corpus, 
the authors of the published paper annotated---in the camera ready version of the paper---the spans that answer the review question or marked the question as unanswerable from the camera ready paper. 

\paragraph{Analysis} To show the need for augmented reading for papers, we collected from OpenReview the versions of the papers submitted for review. We were able to find the submission version for 116 of the PeerQA papers with author answers.

For each answer span annotated in the camera ready version, we search for a corresponding span in the submitted version of the paper, looking for exact text overlap. We find that 52.1\% of the 261 questions marked as answered in the camera ready are also answered in the version submitted for review. The percentage of all PeerQA questions (362) from these papers that can also be answered in the submitted version is 37.6\%.

LinkNav functionality would have surfaced some of the connections that reviewers missed. Reviewing is one setting in which the reader is doing a close reading of the paper and the omissions are most likely an indication of the complexity of the academic paper genre. Explicit linking and surfacing of content can help skimming starting from the abstract or closer reading as in reviewing.

\section{Putting LinkNav Together}
First, an article is split into approximately 512 token segments, adjusted for paragraph and section boundaries and the presence of figures or tables. Figures and tables are considered their own paragraphs, and segments are built by continually adding paragraphs until adding the next paragraph would exceed the 512 token limit. We also ensure that at section boundaries a new segment is created to prevent text from multiple sections appearing in a single segment.

LinkNav functionality is supported by three components as shown in Figure~\ref{fig:LinkNavPipeline}: \emph{(a)} question generation of plausible questions that may arise while reading a segment; \emph{(b)} answer detection to find if an answer to that question exists in the paper. Detection is done by computing similarity between a question and paper segments. The top most similar segments and the question are then passed to a language model, which decides if any of these segments answer the question; \emph{(c)} create a directed graph structure that links segments $i$ and $j$ if a question that is generated for segment $i$ is answered in segment $j$. Questions for which an answer is not found in the paper are discarded and do not result in a link between segments. 
Each component can be instantiated in a more sophisticated manner, likely leading to improved performance but certainly more costly.

Here we evaluate if the solution with fairly 
cost-effective
components can be deemed sufficient for a minimal viable experience. For evaluation we use a randomly sampled set of 29 papers from the PeerQA papers with author annotated question answers. We call this subset PeerQA-Gold because of the availability of highly accurate gold answers. This subset is large enough to provide meaningful measurements but not unreasonably costly to compute the results.

\subsection{Question Generation}
\label{sec:q_gen}

To generate questions, we pass each segment to \texttt{\small llama-4-maverick-17B-128E-Instruct} with the prompt shown in the Appendix~\ref{sec:qgen_prompt}, to obtain meaningful questions that can arise while reading that segment. Prior work has shown that the vast majority of questions generated in this manner are valid, reasonable questions \cite{wu-etal-2023-qudeval,wu-etal-2024-questions}.

\begin{table}[h]
    \centering
    \small
\begin{tabular}{lrrrr}
\toprule
Model      & \#Qs & \% Ans  & Segment Qs & Distance \\
\midrule
GPT-4o     & 4583 & 61.292 & 5.348 & 11.503 \\
o3         & 4741 & 32.272 & 5.532 & 11.078 \\
Llama-4    & 3558 & 79.174 & 4.152 & 10.491 \\
\bottomrule
\end{tabular}
    \caption{Number of questions generated (\#Qs), percent of answerable questions (\% Ans), average number of questions generated for a document segment (Segment Qs), and average distance in segments between connecting segments for valid connections (Distance) for three different question-generation models.}
    \label{tab:answer_stats}
\end{table}

Table~\ref{tab:answer_stats} shows statistics for the questions generated for the PeerQA-Gold subset papers with three models: two closed-source models, \texttt{gpt-4o} and \texttt{o3}, and one open-source model \texttt{llama-4}. 
For this analysis only, we use a slightly different heuristic for splitting documents into segments, only taking into account section boundaries and the 512 token limit to ensure a more uniform set of segments.
The models generate a different number of questions, with \texttt{Llama-4} producing the fewest questions. However \texttt{Llama-4} generates the largest number questions that are answered in the paper; about 80\% of  the questions it generates are answered somewhere in the document. In contrast, \texttt{o3} generates many more questions, but only over a third of them are answered in the paper itself. Despite these differences, ultimately all three models generate five connections per segment via questions that are answerable later in the paper. The segments where the question arises and the segment that contains the answer are 10 segments apart from each other on average. 

Given the similar distribution of questions per segment and the distance between connected segments for all three models, we choose to use \texttt{Llama-4}. \texttt{Llama} creates a similar number of connections as the other models but with considerably fewer questions for which answers have to be found in the document. The LinkNav experience with \texttt{Llama-4} is thus faster to create and cheaper than using the other models.

\begin{table}[h]
\centering
\small
\begin{tabular}{lccc}
\toprule
System Pair (A, B) & A Unique & B Unique & Common \\
\midrule
(GPT-4o, o3)      & 53.67 & 28.70 & 17.63 \\
(GPT-4o, Llama-4) & 41.37 & 34.54 & 24.09 \\
(o3, Llama-4)     & 33.15 & 51.52 & 15.33 \\
\bottomrule
\end{tabular}
\caption{For each model pair (A, B), the percentage of connections made by A only (A Unique), by B only (B Unique), and by both systems (Common) out of the union of connections, averaged per paper.}
\label{tab:unique_connections}
\end{table}

Although all three models produce a similar number of connections and comparable distances between linked segments, they often connect different segments. Table~\ref{tab:unique_connections} shows the pairwise overlap in connections for the PeerQA-Gold papers. 
Across all pairs, the majority of connections made by each system are \emph{not} shared with the other, with unique connections consistently outweighing common ones.
Evaluating human preference for specific connections is beyond the scope of this work, but these substantial differences warrant a future comparison of perceived differences in the resulting experience.

\subsection{Answer Detection}
\label{sec:ans_dect}

After the questions are generated, answer detection finds if an answer to each question exists in the paper. To keep runtime and cost reasonable, detection is done by computing similarity between the question and paper segments. The top most similar segments and the question are then passed to a language model, which decides if any of these segments answer the question. 

Specifically, we use the OpenAI \texttt{text-embedding-small} embeddings to find the $n$ paper segments that are most similar to the question. We seek the smallest $n$ that would ensure that the correct answer is among the top $n$, to minimize the length of text passed on to the model making the final decision if the question is answered, while ensuring that most questions with an answer are answered. 

\begin{figure}
    \centering
    \includegraphics[width=1\linewidth]{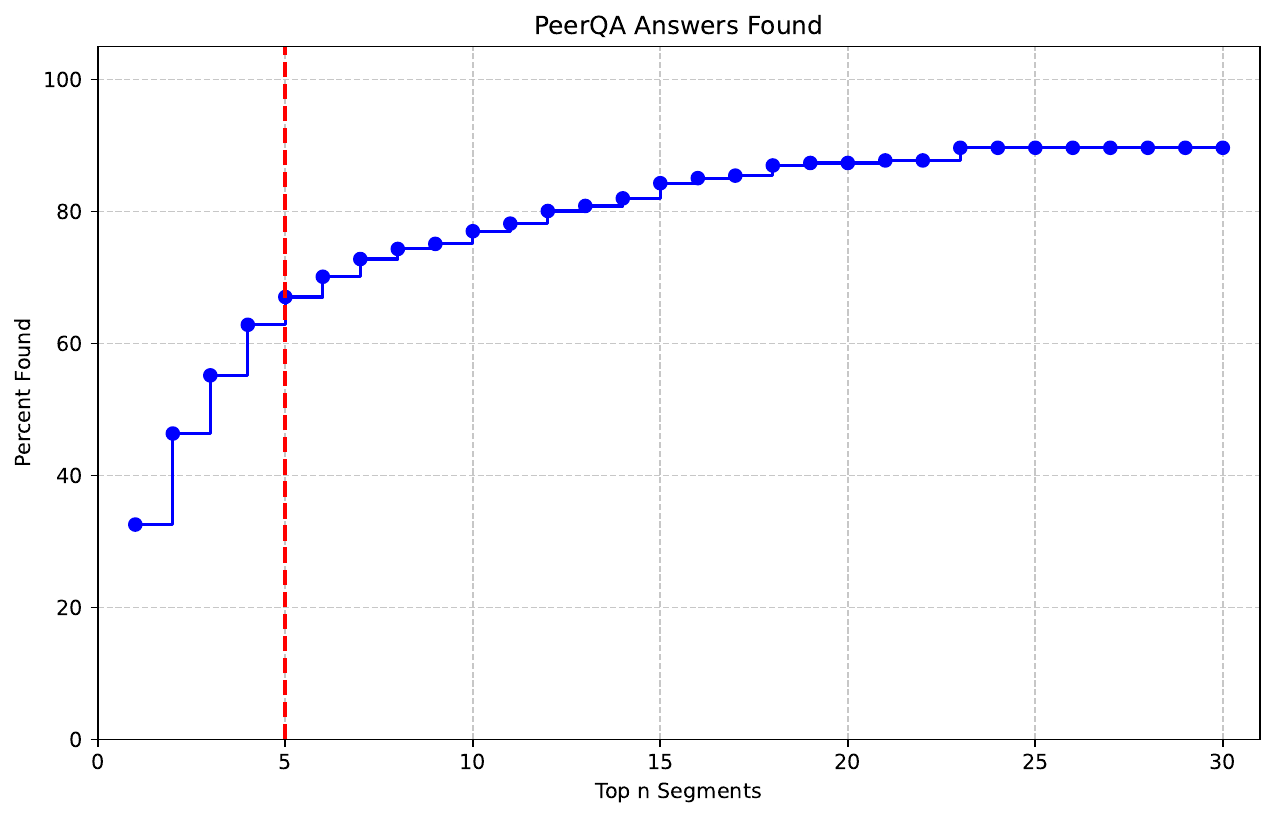}
    \caption{Cumulative percent of PeerQA-Gold answers in the Top $n$ segments most similar to the question. Beyond the top-5 segments, the growth in this cumulative percent slows down significantly.}
    \label{fig:cdf}
\end{figure}

Figure~\ref{fig:cdf} shows the cummulatative distribution function for the percent of answerable questions in the PeerQA-Gold dataset. For the first five steps adding an extra segment increases the coverage of true answers by over 5\% at each steps. After that, the growth slows down, with 95\% of answers in the top 30 segments. We choose to pass on the top 5 segments to the model for final decision. At this cut-off, 72\% of questions with known gold annotated answer have a segment answer in the top 5. Like this, we lose some connections in the text, trading these off for speed and cost.

After that we use a language model to decide if any of the five segments answers the question. If none of the segments is an answer to the question, the question is not suitable for content linking. If a segment is identified as answering the question, the segment where the question arose and the answer segment are connected in LinkNav.

In table \ref{tab:f1_results} we present results for \texttt{gpt-5-mini} and \texttt{gpt-4o} on PeerQA-Gold for  finding an answer to a question $q$ among the five segments most similar to $q$. Precision indicates how often a segment identified as an answer to the question is indeed an answer and recall indicates what percentage of existing answers are identified.\footnote{For 28\% of the questions the answer segment is not among the top five as seen in Figure~\ref{fig:cdf}, lowering the maximum obtainable recall. Some questions have more than one segment that answer the question, so recall can exceed 72\%.}   
We present the prompt in Appendix~\ref{app:answer_prompt}.


\begin{table}[h]
    \centering
    \small
    \begin{tabular}{lcccc}
        \toprule
        Model & Precision & Recall & F1 \\
        \midrule
        \texttt{gpt-5-mini} & 0.798 & 0.801 & 0.799 \\
        \texttt{gpt-4o} & 0.836 & 0.586 & 0.689 \\
        \bottomrule
    \end{tabular}
    \caption{Precision, recall, and F1 @ 5 (among the top 5 answers) for models deciding if a segment is an answer to a given question. Evaluated against ground truth PeerQA-Gold.}
    \label{tab:f1_results}
\end{table}

The precision of 4o is about 5\% higher. 
Precision is important for the LinkNav experience because if two segments are connected in the interface but the connected segment does not answer the question that triggered the link,  
the experience will not be intuitive. With 80\% precision, most of the connections will be well-justified and either model can be used. For recall however,  \texttt{gpt-5-mini} is 20\% absolute difference better than \texttt{gpt-4o}.  \texttt{gpt-4o} will miss 40\% of connections between segments, leading to an incomplete experience. \texttt{gpt-5-mini} is cheaper, faster and overall more suitable for use in LinkNav.

\section{Properties of Connections}


\begin{table}[h]
    \centering
    \small
\begin{tabular}{lrrrrrrrrr}
\toprule
Model & Before & After & Same & Connect \\
\midrule
GPT-4o & 27.67 & 7.53 & 30.15 & 1369 \\
o3 & 27.78 & 7.52 & 26.24 & 848 \\
Llama-4 & 28.93 & 7.22 & 33.60 & 1146 \\
\bottomrule
\end{tabular}
    \caption{Percent of unfiltered question-answer connections connecting backwards (Before), percent of unfiltered question-answer connections connecting to immediately forward adjacent segments (After), percent of unfiltered connections connecting to the same segment as the question (Same), and number of valid connections after filtering (Connect).}
    \label{tab:big_table}
\end{table}
\begin{table}[h]
    \centering
    \small
    \begin{tabular}{llrrrrr}
    \toprule
     System & $d$ = 0 & $d$ = 1 & $d$ = 2 & $d$ = 3 & $d$ > 3 \\
    \midrule
GPT-4o & 45.040 & 19.837 & 12.369 & 7.468 & 15.286 \\
o3 & 52.509 & 23.337 & 11.552 & 5.484 & 7.118 \\
Llama-4 & 46.558 & 21.937 & 10.268 & 9.802 & 11.435 \\
    \bottomrule
    \end{tabular}
    \caption{Percent of all segments corresponding to their in-degree $d$ for all three systems evaluated.}
    \label{tab:indegree}
\end{table}

Here, we present additional statistics about the segment connections established via each of the three models, \texttt{gpt-4o}, \texttt{o3} and \texttt{llama-4}.

Answers that appear before or in near proximity to the question origin are presumably low utility, as the reader have either read or will soon read the information that will address that question. So all question-answer connections where the answer segment preceded, coincided with, or immediately followed the question's origin segment are filtered out in the current implementation. Ideally in future work instead we will have the ability to generate only valid questions that link further segments following the point of reading.

The percentage of filtered connections by type is shown in Table~\ref{tab:big_table}. The three models have roughly similar rate of questions leading to invalid connections between segments. About one third of all questions are answered in a segment before the segment where they arise. Presumably the authors organized their writing to have this effect, so details are already shared at the point where the reader needs them. In future work, these planned connections can also be leveraged in an augmented experience, with an info card that summarizes previously shared content.  Questions however are not a good vehicle for introducing the reminder.

Another 30\% of questions are answered in the segment for which they were generated. Augmented reading for these is clearly unnecessary. \texttt{o3} has the fewest (25\%) of these types of questions.

Another 7\% of connections are to segments immediately following the one where the question arose, again suggesting minimal utility at best for making the connection explicit in LinkNav. 

Overall, about 65\% of generated questions are useless, requiring generation and answer validation time and cost but not leading to a meaningful connection. More efficient methods for question generation will be valuable in making the experience practical in the future.

\paragraph{Question Yield}

The number of questions and the number of connections
that could be extracted from these questions are useful factors to consider for document-linking question generation. For each question generated, there is an additional cost in detecting its answer within a document. 
However, the questions generated should also yield utility by revealing a connection. Thus, a higher proportions of answerable questions are desirable. These factors are what led us to choose \texttt{llama-4-maverick-17B-128E-Instruct} as the question generation model in LinkNav. 

\paragraph{In-degree Analysis}

A segment's in-degree $d$ is equal to the number of different questions answered by information in that segment.
 High in-degree indicates that a segment contains important information relevant to content presented throughout the paper. Pre-neural extractive summarization methods have highlighted that a text node with highest in-degree usually contains the most important information in the text \cite{mihalcea-tarau-2004-textrank,10.5555/1622487.1622501}. For the present LinkNav experience however repeated connections to the same content may be distracting if the reader had already read the central segment via a previously created link.

One way to deal with repeated links is to make the text links fade in color once a segment has been visited during reading. Like this, the reader will be aware that they have already seen the information. A cruder way is to attempt to minimize links to the same segment, even if they naturally exist in a text. 

In Table~\ref{tab:indegree}, we present the percent of segments where $d = 0, 1, 2, 3$ and $d > 3$ for each source of questions. Questions generated with o3 leave over half of the segments in the paper without connections, an in-degree of zero. GPT-4o and Llama-4 have much smaller number of unconnected segments.
Out of all evaluated models, \texttt{llama-4} resulted in the highest proportion segments with in-degrees ranging from 1 to 3. This result combined with the factors mentioned earlier motivated our decision to use it as the question generation model for LinkNav.

\section{Conclusion}

Academic papers encode rich semantic connections within a linear structure, often leaving readers to bridge distant passages on their own. We introduce LinkNav, which surfaces such non-adjacent connections by generating reader-aligned questions and identifying answer-bearing segments through dense retrieval and model-based verification. Our evaluation shows that many linked passages are substantially far from each other.
By combining localized answers with direct navigation, LinkNav augments the traditional paper with a semantic layer that supports more coherent and efficient reading.

\section*{Limitations}

Our work deals with the problem of surfacing distant semantic connections between segments of a paper. Readers however may need additional types of support, such as identifying key segments and their interpretation and reminders for content they have already read. The analysis of the paper graph shows that it may support further augmentations but these are beyond the scope of this work. 

In our analysis of LLM-generated questions for intra-document connections, we used one-fourth of the papers we extracted from the PeerQA dataset. This was done mostly for cost and time reasons. The smaller dataset is sufficient to meaningfully measure common events such as number of questions generated per segment, answer recall when using question-segment similarity  and precision and recall for the final answer detection.

Our question generation and subsequent answer detection approaches results in possibly unanswerable questions and low-utility connections that have to be filtered out later. The choice of the model we used for question generation was predicated on minimizing this waste. This paper is centered on this application of surfacing intra-document connections to users via inquisitive questions. Future work will focus on approaches for further optimizing question generation for such connections.

As with any user experience, there is the potential for certain subsets of readers to find low utility in this application. We highlight clear benefits for readers using LinkNav, and future work will focus on user studies to understand readers preferences for surfacing connections in academic papers.

\section*{Ethical Concerns}

Large language models are transforming society, for some making human reading unnecessary. We believe that human competency even in the future will require human reading and seek ways to modernize the experience for doing so.

We use evidence from peer review that reviewers miss some connections and look for information that the authors have already provided in 30\% of their questions. This analysis is not a critique of reviewers, but an evidence that reading is complex and has to be modernized with the emerging capabilities offered by large language models. 

\section*{Acknowledgements} We thank Jack Wang and Alexa Siu for their valuable feedback and comments in the early stages of this work.


\bibliography{custom}

\appendix

\onecolumn

\section{Question Generation Prompt}
\label{sec:qgen_prompt}
\begin{Verbatim}[
    frame=single, 
    framesep=3mm, 
    label=Prompt A, 
    fontfamily=courier, 
    fontsize=\tiny,
    breaklines=true,
    breakanywhere=true,
    breaksymbol={},
    breaksymbolleft=,  % Removes the left return symbol
    breaksymbolright=  % Removes the right return symbol
]

System: You are logical, intelligent, insightful, precise, and can understand the contents of research papers. You are knowledgable on different fields and domains of science and engineering. You are able to interpret research papers, create questions and answers, and compare multiple aspects.

Imagine the following scenario: You are reviewing a submitted research paper manuscript. After you are done reading, you have questions you want to ask the authors. You are a very intelligent reader so you don't ask question that are already answered within the paper.

I will provide you two things: First, context which you have already read, and a chunk of text that you are currently reading.

This is your task. First, you need to look at the context that is provided to you. This is information that you have already read and aware of. Please don't ask questions that can be answered through this context. What you need to do instead is generate all possible follow-up questions arising from the provided chunk of text that you are currently reading. These follow-up questions must be insightful and highlight a critical gap of information that a reader would desire to know while reading.

You MUST follow these rules when creating these questions:
(1) The question must not be answered by text in the context. Make sure to thoroughly read the text in the context and craft a question that cannot be answered by it.
(2) These questions must be questions that a smart reader is naturally thinking about when reading this chunk of text. Remember, they also have already read the context as well, so they will not be thinking of questions that they already know the answer to. Put yourself in this smart reader's shoes when crafting these questions.

I will provide you with a typology of these questions below. You can use this typology to come up with more diverse questions.

Typology:
  - Content: 
    - content-clarification: questions that try to clarify something that you've read
  - Insights:
      - insights-more: questions that ask for more insights
      - insights-nuance: questions that ask for more nuance in the insights
      - insights-soundness: questions that ask whether the insight is sound.
  - Measurement:
      - measurement-more: questions that ask for more metrics to evaluate in experiments
      - measurement-detail: questions that ask for more detail about some measurement/metrics
      - measurement-alternative: questions that ask why an alternative measurement wasn't used
  - Method: 
      - method-alternative: questions that probe at an alternative method that has a similar effect
      - method-detail: questions that probe for more detail about the method
      - method-motivation: questions that probe at the motivation for using such a method

When you respond, first think really hard, step-by-step, about every span of text in the provided chunk, and think about what would be good, thoughtful, and interesting follow-up questions according to the above rules.

Then you need to provide me with these follow-up questions and also tell me the exact span of text each follow-up question corresponds to, as in follows up on, in your response. Absolutely make sure that the span you provide is an exact substring of the chunk text, and that it is not a summary or paraphrase. It should be the exact text corresponding to the question.

If you can't come up with any follow-up questions at all (which is perfectly fine), just return None.

### REASONING PROCESS
  <Explain your reasoning and thinking process>

### JSON
  { 
      "follow_up_questions": [
                          {
                              "question": "<generated follow-up question>",
                              "span": "<exact span of text in the provided chunk corresponding to this question>"
                              "question_type": <Content/Insights/Measurement/Method>,
                              "question_subtype": "<content-clarification/insights-more/insights-nuance/insights-soundness/measurement-more/measurement-detail/measurement-alternative/method-alternative/method-detail/method-motivation>"
                          }, 
                          
                          ...
                      ] or None
  }

User: 

[Start of Context]
{context}
[End of Context]

[[Start of *Provided Chunk*]]
{segment}
[[End of *Provided Chunk*]]

\end{Verbatim}
\newpage

\section{Answerability Prompt}
\label{app:answer_prompt}

\begin{Verbatim}[
    frame=single, 
    framesep=3mm, 
    label=Prompt B, 
    fontfamily=courier, 
    fontsize=\tiny,
    breaklines=true,
    breakanywhere=true,
    breaksymbol={},
    breaksymbolleft=,  % Removes the left return symbol
    breaksymbolright=  % Removes the right return symbol
]

System: You are logical, intelligent, precise, and can understand the contents of research papers. You are knowledgable on different fields and domains of science and engineering. You are able to interpret research papers, and determine whether information inside these papers can answer some question.

This is your task: Read the following several chunks from a paper and answer whether the question can be answered by one or more of these chunks. If the question can be answered, report the chunk numbers of the chunks where you found the answer and answer the question in a sentence or two using only the information in that chunk(s).

When you respond, first think really hard, step-by-step, about every sentence in every chunk, and analyze whether this information can answer the question provided. 

You can definitely find the question to be unanswerable. You can only find a question to be answerable if ONLY the information within the provided chunks fully answers this question.

If you find the question to be answerable, please select the fewest possible chunks you would need to answer the question. If there is one chunk that can fully answer this question, please select only that chunk. If there are multiple chunks that can fully answer this question, please select only the one chunk you think best answers the question. You should only select multiple chunks if the only way to fully answer the question is by combining the information in these chunks.

The chunk numbers are 1-indexed, meaning the first chunk is chunk number 1, the second chunk is chunk number 2, and so on. When you recieve the chunks, you will know the chunk number as it appears like this: "CHUNK #<chunk_number>". Please only report the exact number, and only the number, of the chunks you selected. Do not report a chunk number that is not in the range of possible chunk numbers. The chunk number **hould not be less than 1** nor should it be greater than the maximum chunk number. This is unforgivable. 

In addition to selecting the chunks, you also need to provide the exact span of text within the chunk that answers the question. This span should be a substring of the chunk text that directly addresses the question. Absolutely make sure that the span you provide is an exact substring of the chunk text, and that it is not a summary or paraphrase. It should be the exact text that answers the question.

You need to provide me with an binary Yes or No as the answer to whether the question is answerable by the provided chunks. If answerable, provide me with the list of chunk numbers corresponding to your selected chunks. I also require you detail the exact span of text within the chunk that answers the question. If unanswerable, this can be an empty list.

If answerable, provide me with a 1-2 sentence answer. Otherwise, just answer with an empty string.


### REASONING PROCESS
<Explain your reasoning and thinking process>

### JSON
{
  "is_answerable": "<Yes/No>",
  "selected_chunks": [
    {
      "chunk_number": <chunk number>,
      "span": "<exact span of text within the chunk that answers the question>"
    },
    ...
  ],
  "answer": "<1-2 sentence answer, or empty string if unanswerable>"
}

User: 

[Start of Provided Chunks]
{answer_batch}
[End of Provided Chunks]

[Start of Question]
{question}
[End of Question]

\end{Verbatim}

\section{Hyperparameter Details}

We present details on the specific hyperparameters used here. For OpenAI models (\texttt{gpt-4o}, \texttt{text-embedding-small}, \texttt{gpt-5-mini}, \texttt{o3}), we use default hyperparameters as specified by the OpenAI API. For \texttt{llama-4-maverick-17B-128E-Instruct}, we specify a temperature of 1 and the maximum token produced limited to 4000 tokens.

\end{document}